\documentclass{iau} 
\usepackage{graphicx}

\title[JD 11.~~Central Stars of Planetary Nebulae in Galactic Open Clusters] 
{Central Stars of Planetary Nebulae in Galactic Open Clusters: Providing additional data for the White Dwarf Initial-to-Final-Mass Relation}

\author[Vasiliki Fragkou, Quentin A. Parker, Albert Zijlstra, Richard Shaw, Foteini Lykou]   
{Vasiliki Fragkou$^{1}$,
 Quentin A. Parker$^{1}$,
Albert Zijlstra$^{2}$,
Richard Shaw$^3$,
Foteini Lykou$^{1}$}

\affiliation{$^1$The University of Hong Kong, Department of Physics, Hong Kong SAR, China
 \\ email: {\tt vfrag@hku.hk, quentinp@hku.hk, lykoufc@hku.hk} \\[\affilskip]
$^2$The University of Manchester, Manchester, UK
 \\email: {\tt a.zijlstra@manchester.ac.uk} \\[\affilskip]
$^3$ Space Telescope Science Institute, Maryland, USA
 \\email: {\tt shaw@stsci.edu}
}

\pubyear{2018}
\volume{343}  
\jname{Why Galaxies Care About AGB Stars: A Continuing Challenge through Cosmic Time}
\editors{F. Kerschbaum, M. Groenewegen \& H. Olofsson.}
\begin{document}

\maketitle

\begin{abstract}
Accurate ($<$10\%) distances of Galactic star clusters allow precise estimation of the physical parameters of any physically associated Planetary Nebula (PN) and also that of its central star (CSPN) and its progenitor. The progenitor's mass can be related to the PN's chemical characteristics and furthermore, provides additional data for the widely used white dwarf (WD) initial-to-final mass relation (IFMR) that is crucial for tracing the development of both carbon and nitrogen in entire galaxies. To date there is only one PN (PHR1315- 6555) confirmed to be physically associated with a Galactic open cluster (ESO 96 -SC04) that has a turn-off mass $\sim$2 M$_{\odot}$. Our deep HST photometry was used for the search of the CSPN of this currently unique PN. In this work, we present our results.

\keywords{Planetray Nebulae, Open Clusters}
\end{abstract}

\firstsection 
\section{Introduction}

CSPNe masses, crucial in understanding post-AGB evolution, provide additional data for the widely used WD IFMR. CSPNe studies are difficult as measurements of their masses require precise determination of their distances. PNe members of Galactic star clusters allow the accurate determination of their distances from cluster Color-Magnitude Diagrams (CMDs), while photometric measurements of their CSPNe can constrain their intrinsic luminosity and mass and thus, these objects can be used as additional points for the IFMR. PHR 1315-6555 is a faint, bipolar, possible Type I PN, whose radial-velocity match this of the distant Galactic open cluster ESO 96-SC04 (\cite[Parker et al. 2011]{par11}). 

In this work we present our deep HST WFC3 F555W, F814W, F200LP and F350LP photometry of the cluster, the PN and its CSPN. Following, we create an improved cluster CMD and determine the physical properties of the nebular CSPN.

\section{Methods and Results}

The F555W and F814W HST filter exposures provide the deepest CMD of the cluster available to date. The theoretical Padova isochrone (\cite[Girardi et al. 2000]{gir00}) that best fits our CMD suggests that the distance to the cluster is 10.5 $\pm$ 0.4 kpc, it has an age of 0.72 $\pm$ 0.1 Gyrs, a reddening of 0.695 $\pm$ 0.06 and a turn-off mass of $\sim$ 2 M$_{\odot}$.

\begin{figure}
\begin{center}
 \includegraphics[width=4in]{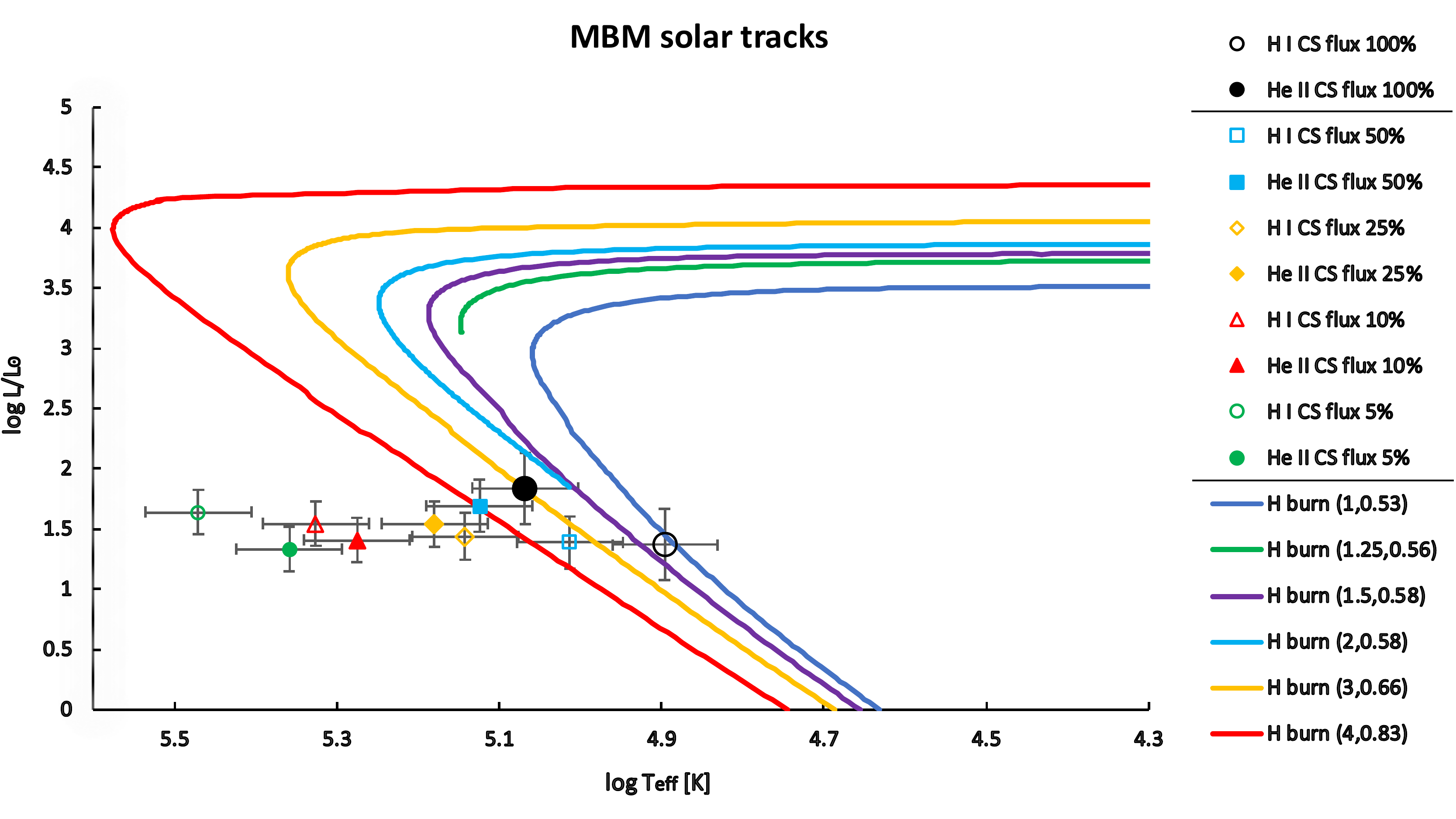} 
 \caption{ Derived CSPN temperatures and luminisities plotted along the MBM tracks.}
   \label{fig1}
\end{center}
\end{figure}

The deep F200LP and F350LP HST filter exposures were used to find the CSPN since the F200LP-F350LP gives the near-UV continuum. This was the bluest star in the nebular field lying close to the projected nebula centre as might be expected.The CSPN temperature and luminosity were derived from the F555W CSPN measured magnitude using the Zanstra method (\cite[Zanstra 1931]{zan31}) and the nebular HI and HeII absolute fluxes (\cite[Parker et al. 2011]{par11}). Previous steps were repeated assuming that the presence of an unresolved companion contributes to the measured CSPN visual flux in fractions from 50\% to 95\%. For all previous cases the CSPN masses were determined by plotting the derived temperatures and luminosities in the logT-logL plane along with the \cite [Miller Berolami] {mbm} (2016, MBM) post-AGB evolutionary tracks for solar metallicities (see Fig. \,\ref{fig1}). The CSPN mass was found to be 0.65 $\pm$ 0.09 and the presence of a companion contributing to the measured flux $>$ 50\% is deemed highly unlikely as this would place the CSPN too far from the evolutionary track expected from the estimated turn-off mass of the cluster. The  measured F814W magnitude can provide any indications regarding the possible presence of an unresolved binary cool low mass companion. This would yield an IR excess (see \cite [Barker et al. 2018]{bar18}). $(F555W-F814W)_{obs}$ - $(F555W-F814W)_{mod}$ = 0.013 $\pm$ 0.67 and thus, the presence of a binary cool companion is not favored by the data.          

The nebular log(N/O) abundance ratio has been measured to be around 0.87 (Parker et al. 2011). Stellar evolution models predict that such high N abundances in solar metallicity environments would be for stars with initial mass $>$ 4 $M_\odot$ (\cite[Karakas \& Lugaro 2016] {kar16}). which is not the case here.  A much lower metallicity environment could explain the N yields but is not supported from our CMD. If cluster metallicity is proven to be close to solar the lower mass limits for hot bottom burning may need to be revised given our results. 

\end{document}